\documentclass[11pt,twoside]{article}
\usepackage{asp2004}
\usepackage{psfig}
\usepackage{epsf}
\usepackage{graphics}
\usepackage{lscape}
\def\edcomment#1{\iffalse\marginpar{\raggedright\sl#1\/}\else\relax\fi}
\reversemarginpar

\begin{document}
\title{Binaries as Astrophysical Laboratories: Open Questions}
 \author{I. Ribas}
\affil{Institut de Ci\`encies de l'Espai -- CSIC, Campus UAB, Facultat de
Ci\`encies, Torre C5 - parell - 2a planta, 08193 Bellaterra, Spain; e-mail:
iribas@ieec.uab.es}
\affil{Institut d'Estudis Espacials de Catalunya (IEEC), Edif. Nexus, C/Gran
Capit\`a, 2-4, 08034 Barcelona, Spain}

\begin{abstract}
Binary systems have long been recognized as the source of powerful
astrophysical diagnostics. Among the many applications of binary stars,
they have been used as probes of stellar structure and evolution (both of
single and binary stars) in a broad range of masses, evolutionary stages,
and chemical compositions, and as indicators of distance and time. With
the numerous ongoing photometric surveys and upcoming space astrometry and
photometry missions, the future of binaries looks bright. The various
aspects of binaries as astrophysically useful laboratories are reviewed
here, with emphasis on the currently open problems and research
opportunities. 
\end{abstract}

\vspace{-0.5cm}
\section{Introduction}

Any star that belongs to a binary system becomes automatically more
valuable. There are several characteristics that make stars orbiting each
other especially noteworthy. For example, the orbital motion of a binary
system makes it possible to directly determine the fundamental properties
of the component stars to high accuracy. Stellar masses can be determined
from radial velocities and/or direct astrometric measurements.
Furthermore, if the binary system happens to have an orbital inclination
close to 90 degrees, the components undergo mutual eclipses and the
resulting light curve yields direct measurements of the stellar radii. The
high-precision stellar properties from the analysis of binary stars
constitute a very useful dataset to carry out stringent tests of stellar
models. But also binaries have been successfully used as indicators of
distance and time.

Binaries themselves are very interesting subjects of study. For
example, tidal interactions or mass accretion alter the orbital properties
and the evolution of stars in binary systems and provide valuable insight
into the physical laws that govern those processes. In addition, binary
systems can be associated with energetic phenomena such as cataclysmic
binaries and X-ray binaries, and are the progenitors of objects of strong
astrophysical interest such as novae, supernovae, gamma-ray bursters, etc.  
Calculations also indicate that double degenerate systems will be strong
sources of gravitational waves, which is a new area of research that is
bound to acquire great relevance.

In the era of surveys, binary systems with photometric variability
(especially eclipsing binaries) are being reported by the thousands. But
even larger numbers (millions) are expected in the next decade. Here I
review some of the applications of binaries as astrophysical laboratories
with special focus on the questions that remain open in the different
subjects. These open questions can equally be regarded as opportunities
for research with well defined objectives.

\section{Binaries as probes of stellar structure and evolution}

The aim of stellar structure and evolution models is to produce a
physically sound description of the interior of stars and thus a realistic
picture of their evolution (as a function of the initial mass and chemical
composition). Obviously, a direct view of the stellar interior is very
difficult to obtain (except for measurements using asteroseismology) but
theoretical models make predictions about the macroscopic properties of
the stars, such as temperatures, masses, radii, densities, etc, that can
be {\em and need to be} tested against observations. The comparison of
model predictions with observations is more stringent when the number of
free parameters is very small or null. Models will pass the test {\em
only} if they are able to reproduce {\em all} of the observed stellar
properties given the available constraints. Detached eclipsing binary
stars, with their accurate determinations of their absolute dimensions,
provide the best tests of stellar models \citep[see, e.g., the thorough
review by][]{A91}. The detached restriction is set to guarantee that the
components of the binary system have evolved as single stars.

In this section, I review the comparison of binary star data with
theoretical models. To do so, the section has been subdivided to cover
high-mass and low-mass stars in the main sequence, which are subject to
different physical mechanisms and thus have different issues. Also, this
section addresses the use of binaries to study the extended atmospheres of
cool stars (either evolved or in the main sequence), and briefly some of
the open problems in a ``new'' and popular type of binary/multiple
system: planetary systems.

\subsection{High-mass stars}

High-mass stars are very important for many astrophysical processes, e.g.,
emission of ionizing radiation, chemical evolution of the galaxy,
energetic phenomena. Thus, a good characterization of their evolution off
the zero-age main sequence (ZAMS) is of central importance to understand
all subsequent processes. There are several physical mechanisms that
acquire great relevance when modeling the evolution of high-mass stars.
Convection parameters, rotation effects and mass loss are some of them. In
particular, it is worth pointing out that our current theoretical
description of convection is still rather crude and, although there have
been advances in other directions \citep{CM91}, the parametric and
phenomenological mixing-length theory \citep{BV58} is still widely used
\citep*[e.g.,][]{SDG05}. The main two parameters in the mixing-length
theory are the mixing-length parameter, which is usually determined by
comparison between the observed and predicted solar radius, and the
convective overshoot parameter, which is more difficult to assess.

Convective overshoot in the stellar core basically modifies its size and
has the observable effect of expanding the duration of the main sequence
phase. The effect becomes more prominent with increasing mass of the star.  
Overshoot is often parameterized by the value $\alpha_{\rm ov}$, which is
the extension of the core size beyond the Schwarzschild boundary in units
of the pressure scale height. A possible way to estimate the convective
overshoot parameter is the study of the observational color-magnitude
diagram of young clusters. Such studies \citep[e.g.,][]{PD74,MM81,PSH98}
have proved that convective overshoot is relevant and that the location of
the terminal age main sequence of several young clusters is best described
by a convective overshoot parameter $\alpha_{\rm ov}$ of about 0.25.

The high-accuracy stellar fundamental properties of detached eclipsing
binaries were used by \citet*{ACN90} to also place constraints on the
amount of core overshoot. More recently, studies using larger samples of
eclipsing binaries \citep{PTS97,RJT00,LVG02} have confirmed the need for
convective overshoot in the amount of $\alpha_{\rm ov}\sim0.25$ for stars
of intermediate masses (1.5--3~M$_{\odot}$). However, two studies have
further suggested the existence of an increase in the amount of core
overshoot with stellar mass. Using binary systems with one or both
component in an evolved stage (core helium burning phase), \citet*{SPE97}
found a value of $\alpha_{\rm ov}\sim0.24$ for 2.5-M$_{\odot}$ stars
slightly increasing to $\sim0.32$ for 6.5-M$_{\odot}$ stars.
\citet*{RJG00} find that the overshooting parameter may increase up to
$\alpha_{\rm ov}\sim0.6$ for stars of $\sim$10--12 M$_{\odot}$. This
latter result is based on the analysis of the eclipsing binary V380 Cyg by
\citet{GRF00} and it is worth reviewing the main points of the study here.

\begin{figure}[!t]
\plotone{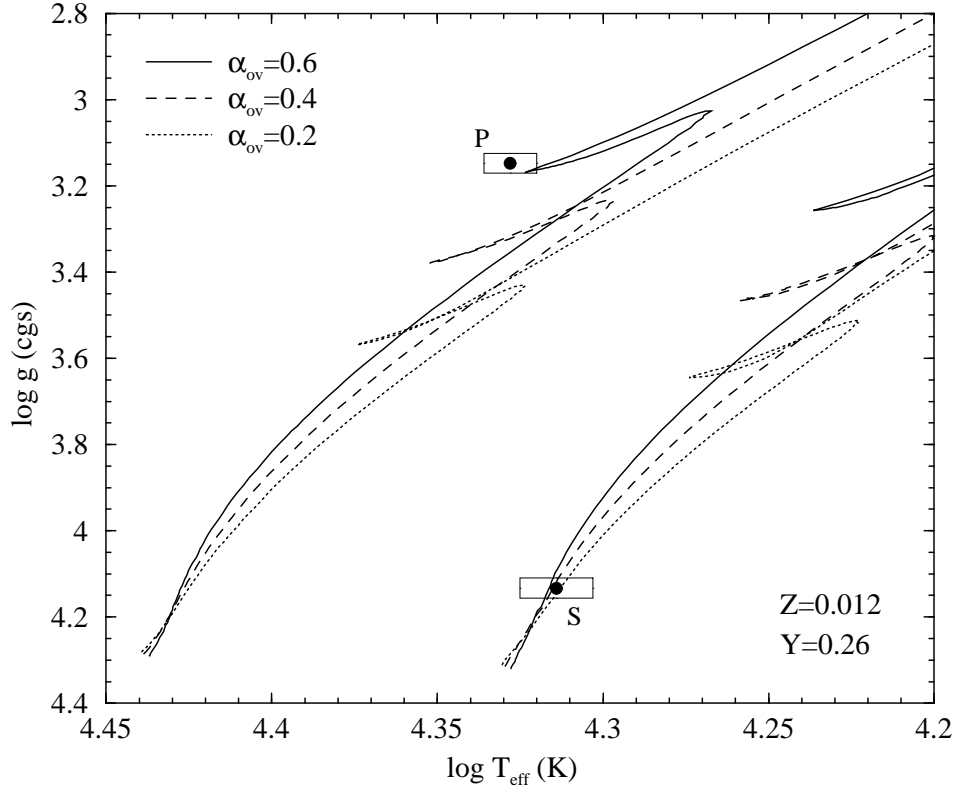}
\caption{$\log g - \log T_{\rm eff}$ plot of V380~Cyg. Evolutionary tracks
for the primary (P) and secondary (S) components computed with
$\alpha_{\rm ov}=0.2$, $0.4$, and $0.6$ are shown.}
\label{fig:v380}
\end{figure}

V380 Cyg may be the prototypical case of a stellar astrophysical
laboratory. This eclipsing system is composed of two B-type stars of
similar temperature but different evolutionary stages. The more massive
component ($M\sim11$~M$_{\odot}$) has a low surface gravity of $\log
g=3.15$ while the secondary component ($M\sim7$~M$_{\odot}$) has barely
left the ZAMS ($\log g=4.13$). The temperatures and metallicity of the
system components could be determined to high accuracy from the fit of the
spectral energy distribution in the UV/optical. The very unequal positions
of the components in the HR diagram makes this system highly discriminant
when testing the performance of evolutionary models. This fact is
illustrated in Fig. \ref{fig:v380}, which depicts a $\log T_{\rm eff} -
\log g$ diagram with the components and evolutionary tracks \citep[kindly
computed by A. Claret using the prescriptions in][]{C95} with different
amounts of convective overshoot. As can be seen, the secondary component's
position is weakly influenced by overshooting and thus fixes the amount of
helium in the models (which is treated as a free parameter). With the
metallicity, helium abundance, $\log g$ and $\log T_{\rm eff}$ of the
primary component known, there are no degrees of freedom left in the
comparison with models other than the value of the overshooting parameter.
In this case, it is concluded that the physical properties of the primary
are only reproduced for a high overshooting parameter of $\alpha_{\rm
ov}\sim0.6$.

But V380 Cyg has the added value of being in an eccentric orbit and the
existence of old timings makes it possible to measure an apsidal motion
period of about 1500 yr. The rate of apsidal motion is directly related to
the internal concentration of the star (i.e., ratio of mean to central
density). Therefore, this provides a further independent check to stellar
models since the amount of convective overshoot is correlated with the
size of the stellar core and thus with the internal concentration of the
star. As shown by \citet{GRF00}, the apsidal motion test also suggests an
overshooting parameter of $\alpha_{\rm ov}\sim0.6$.

A major step forward in the understanding of core overshoot came with the
asteroseismological study of \citet{ATD03}. The analysis of long
time-series photometry of the 9~$M_{\odot}$ star HD 129929 led to the
conclusion that the best match of the models to the observed frequencies
occurs for an overshooting parameter of $\alpha_{\rm ov}=0.10\pm0.05$. How
can the V380 Cyg result and this one, both apparently robust, be
reconciled? No definitive answer is available yet. A possible clue could
be the effect of stellar rotation. The primary component of V380 Cyg has a
rotational velocity of about 100~km~s$^{-1}$ while HD 129929 only rotates
at 2~km~s$^{-1}$. Rotation can have a similar effect to convective
overshoot on the evolution of star in the sense that it alters the
duration of the main sequence phase. The analysis of V380 Cyg is not able
to discriminate between core overshoot and rotation. Thus, the conclusions
can be reformulated to say that the observations indicate a larger
convective core than predicted by the standard models by 0.6 times the
pressure scale height. In view of the results for HD 129929, perhaps the
extra core size has a small contribution from convective overshoot and a
larger one from rotation.

In any case, most of the information on convective overshoot for massive
stars hinges on the analysis of just two stars. High-mass eclipsing
binaries with evolved and unequal components will provide additional clues
to help resolve the current issues and improve our theoretical modelling
efforts.

\subsection{Low-mass stars}

A large fraction of the stars in the Galaxy have masses well below that of
the Sun. In spite of the shear numbers, detailed investigations of the
properties of low-mass stars have been hampered by their intrinsic
faintness. The observation and study of low-mass stars is now experiencing
a rapid development because of the increasing number of deep photometric
surveys and the advent of powerful instrumentation able to obtain
spectroscopy of these faint stars. But also renewed interest arises from
one of the ``hot topics'' of this past decade: exoplanets. Low mass stars,
brown dwarfs, and giant planets share many physical characteristics and
their study and modeling is closely related.

Current stellar structure models of low mass stars have reached a high
level sophistication and maturity \citep[e.g.,][]{CB00,CBA05}. However,
theoretical progress has not been matched by observational developments
because of the difficulty in obtaining accurate determinations of the
physical properties of low-mass stars. The best source of such
high-quality stellar properties comes from the analysis of double-lined
EBs with detached components. For decades only two bona-fide EBs with
M-type components were known: The member of the Castor multiple system YY
Gem \citep{LS78,TR02} and CM Dra \citep{L77,MML96}. Recently,
\citet{DFM99} reported the discovery of eclipses in the CU Cnc and
\citet{R03} carried out accurate determinations of the components'
physical properties. Three additional new M-type EBs have been studied in
detail. These are BW3 V38 \citep{MM04}, TrES-Her0-07621 \citep{Cea05}, and
GU Boo \citep{LR05}. Unfortunately, the quality of the available
observations for BW3 V38 and TrES-Her0-07621 does not permit high-accuracy
determinations of both masses and radii but GU Boo has well-determined
physical properties that make it twin system of YY Gem.

Unfortunately, the number of known low-mass EB systems is still small
because of the faintness of the stars and the often strong intrinsic
variations due to magnetic activity. Another source of potentially
accurate data is the observation of visual binaries and the direct
measurement of the component radii using interferometry. Although there
has been significant progress in this direction -- and more is expected in
the coming years, -- the accuracy reached \citep*{LBK01,SKF03} is not yet
quite sufficient to place stringent constraints on the models. Fundamental
properties of low-mass stars have also resulted from follow-up
observations of OGLE transit candidates \citep{BPM05,PBM05}. However, the
determinations are model-dependent to some extent and the accuracy is
significantly lower than that resulting from double-lined EBs.

\begin{figure}[!t]
\plotone{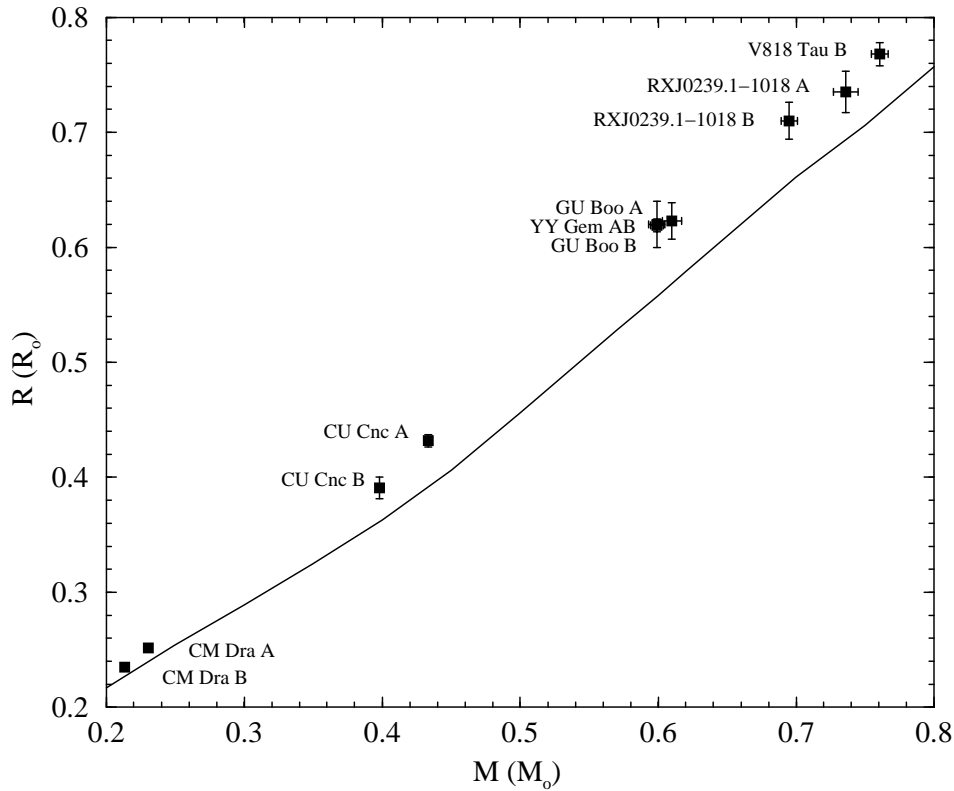}
\caption{Mass-radius plot for the components of EBs in the lower main
sequence with accuracies better than 3\%. The solid line represents a
theoretical 300~Myr isochrone calculated with the \citet{BCA98} models.}
\label{fig:mr}
\end{figure}

The best stellar data from double-lined EBs offer an excellent opportunity
to carry out critical tests to evaluate the performance of low-mass
stellar models. Such tests have been carried out by a number of authors in
the past \citep{P97,CBC99,TR02,R03}, who have systematically pointed out a
(rather serious) discrepancy between the stellar radii predicted by theory
and the observations. Model calculations appear to underestimate stellar
radii by $\sim$10\%, which is a highly significant difference given the
observational uncertainties. This is clearly illustrated in Fig.
\ref{fig:mr}, which shows a mass-radius diagram for M-type EBs with
accurate parameters ($\sigma<3\%$). Also included in the plot are the two
K-type EBs V818 Tau \citep{TR02} and RXJ0239.1-1028 (L\'opez-Morales et
al., in prep.).

In addition to the radius discrepancy, other detailed comparisons have
also shown that the stellar effective temperatures appear to be
overestimated by $\sim$5\%. Complementary, the mass-luminosity plot
\citep{DFS00} seems to be well reproduced by theoretical models
(especially in the K-band, where the effect of starspots is small). All
the evidence together seems to argue in favor of a scenario in which the
stars have larger radius and cooler temperature than predicted by models
but just in the right proportions to yield identical luminosities. The
reason for such apparent coincidence is yet to be understood. A possible
explanation for the discrepancy between models and observations may found
in the effects of stellar activity, which close binaries experience
strongly because they are forced to spin up in orbital synchronism. The
larger radii and lower temperatures could be a reflex of such enhanced
activity. Perhaps a significant spot areal coverage has the effect of
lowering the overall photospheric temperature, which the star compensates
by increasing its radius to conserve the total radiative flux. A more
detailed discussion is provided in \citet{R05} and similar arguments have
been used by \citet{TLM05} to explain discrepancies observed in
higher-mass stars.

The conclusion is that current models may only applicable to inactive
stars, but this is a severe shortcoming since low-mass stars of relatively
young ages are known to be very active. In any case, the discussion above
illustrates that open problems still exist at the most basic levels, i.e.,
even in the description of the masses and radii of stars. More
observations (leading to improved statistics) and further refinements in
the theory of stellar interiors (including the effects of magnetic fields)
will be needed to settle the current issues and achieve a full
understanding of the properties of low mass stars of all ages.

\subsection{Cool star atmospheres}

There are certain evolutionary pathways that lead to binary systems with
components of very unequal temperatures. This is the case of post-common
envelope or post-mass transfer systems with a late-type star and a hot
white dwarf or subdwarf. Especially interesting in terms of their
astrophysical value are eclipsing systems such as the renowned Hyades
binary V471 Tau \citep{NY70} or FF Aqr \citep{DLE77}. There are also
detached eclipsing binary pairs, known as $\zeta$ Aur systems
\citep[e.g.,][]{W70}, composed of a massive star that has evolved into a
cool supergiant and a hot, less massive companion that still remains in
the main sequence.

The study of these stars, with components in very different evolutionary
stages and with large contrasts in temperature and radius, provides
valuable information on stellar evolution and mass transfer. However, most
of the interest in these systems has been driven by the possibility of
using the hot component as a probe of the atmosphere and circumstellar
environment of the cool companion. For example, International Ultraviolet
Explorer spectra of the eclipse ingress and egress phases in V471 Tau were
used by \citet{GWB86} to detect prominence-like structures in the
atmosphere of the K-type component. In the case of $\zeta$ Aur systems, an
illustration of their use to address questions related to mass loss in
supergiant stars was provided by \citet*{CHR83}. Although it may seem that
continued efforts for over two decades should have resolved all lingering
issues, this is actually not the case. As discussed very recently by
\citet{HBB05}, many aspects of the mechanisms responsible for mass loss in
evolved supergiant stars are still poorly known, such as for example, the
wind acceleration. Further studies of $\zeta$ Aur binary systems, in the
UV, optical and radio domains, should help to shed new light on the
currently open questions.

\subsection{Planetary systems}

Much interest has raised the discovery of exoplanets during the past
decade. The quest for new planetary systems beyond our own is so appealing
and has such social impact that is becoming one of the major goals of
national funding agencies and the driving force of a large community of
scientists. But seen in perspective, this ``new'' field is not much
different from the ``classical'' binary studies in the sense that it uses
the same techniques (radial velocity curves and light curves) with
improved precision. Thus, in a broad sense, planetary systems are just
particular cases of binary or multiple systems with components of very
unequal masses.

Using radial velocity and transit techniques, different groups have now
reported some 170 planets (see the updated list at {\tt
www.obspm.fr/planets}). Much can be learned about planetary formation and
evolution from the analysis of the distribution of planets as a function
of different orbital and physical parameters with the increasing
statistical significance of the sample \citep[e.g.,][]{MBF05}. For
example, new concepts such as orbital migration have emerged in recent
years to explain the presence of gaseous giant planets at close orbital
distances. But a specially valuable source of information is that of
transiting planets, in which case the actual mass (not just $M \sin i$)
and the actual radius can be measured. A surprise came already with the
first transiting planet reported \citep[HD 209458 b;][]{CBL00,HMB00}. The
measured radius and mass resulted in a planet with a density significantly
lower than that of Jupiter. Many models have been put forward to explain
such large radius: irradiation from the host star \citep[e.g.,][]{CBB04},
core size \citep[e.g.,][]{LWV05}, or tidal heating \citep*{BLM01,BLL03}.
Although the latter explanation has been ruled out from observations of
null orbital eccentricity \citep{DSR05,LMV05}.

\begin{figure}[!t]
\plotone{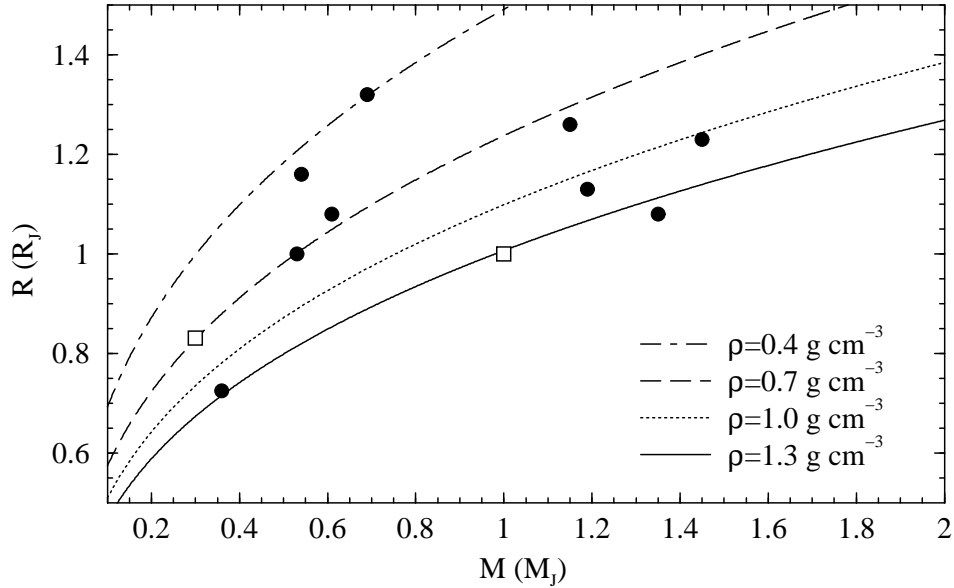}
\caption{Mass-radius diagram of all currently know transiting exoplanets
(filled circles) and the Solar System planets Jupiter and Saturn (open 
squares). Also represented are isodensity lines for various density 
values.}
\label{fig:planets}
\end{figure}

The sample of transiting exoplanets has now increased to 9 and the
mass-radius diagram in Fig. \ref{fig:planets} shows a large variety of
planetary densities (differences of up to a factor of 3). This is surprising
since the planets detected so far constitute a rather homogeneous group, with
similar orbital distances, similar host stars, etc. Different scenarios
involving, irradiation, core sizes or evaporation are currently being
investigated, but such dispersion in the intrinsic properties of otherwise
quite similar planets still defies explanation.

\section{Binaries as distance and time indicators}

Besides providing tests of stellar models, binaries have also been
successfully used as indicators. Most notably, visual binaries and
eclipsing binaries have proved to yield reliable distances potentially
accurate to a few percent. A particularly interesting case is when the
binary system belongs to a larger structure, such as an open cluster or
galaxy, and its distance can be used to estimate that of the host cluster
or galaxy. On the other hand, the strict periodicity of the orbital motion
of binary stars can be used to search for further orbiting companions via
the light-travel time effect, a method much like the one used by O.
R\"omer in 1676 to claim a finite value for the speed of light \citep[see]
[for a complete discussion]{S05}.

\subsection{Distances}

Distances are crucial for a precise knowledge of the scales of the
Universe and thus its structure and ultimate evolution. Since there is no
single distance determination method that can cover from the Solar System
to distant galaxies, the distance scale is built by concatenating a series
of indicators in which each one is used to calibrate the next. Among the
main rungs of this ladder are the Solar System, nearby stars and clusters,
Local Group galaxies, and distant galaxies. Accurate distances to open
clusters and Local Group galaxies are thus of central importance because
the overall cosmic distance scale hinges on them. As shown here, binaries
have made and will make significant contributions to this area.

Distance estimation using binaries can be approached in two distinct ways.
In the case of visual binaries, the so called method of the orbital
parallax is based on comparing the angular size of the orbital semi-major
axis observed from Earth with the true size measured by combining the
elements from the orbital and the radial velocity curve solutions.
Basically, it is equivalent to the classical trigonometric parallax method
except for using the orbit of a binary system instead of the orbit of the
Earth. The other approach is based on the fact that the intrinsic
luminosities of the components of an eclipsing binary system can be
determined directly from the analysis of the light and radial velocity
curves and a temperature calibration. Then, comparison of the observed
brightness with the intrinsic luminosity yields the distance via the
inverse square law. These two distance determination concepts are further
elaborated below.

The calculation of orbital parallaxes of visual binaries is a powerful and
completely direct method to estimate distances \citep[see the review
by][]{Q01}.  The results do not rely on any calibration and thus are
extremely robust. For the method to be applicable, both astrometric data
and spectroscopic data are needed. This is, in fact, its main limitation,
because the radial velocity amplitude decreases with increasing orbital
semi-major axis, and vice-versa. The recent improvements in the accuracy
of the astrometric (using speckle or interferometry) and radial velocity
measurements makes it possible to apply the method to a large number of
visual binaries and not just to a handful of nearby ones. Current
instrumental capabilities yield orbital parallaxes with accuracies $<$1\%
at distances of over 100 pc (to be compared with $\sim$10\% for Hipparcos
trigonometric parallaxes).

The determination of orbital parallaxes by, e.g., \citet{HAB95} and
\citet{KL04} illustrate the capabilities of this method. Another
particularly interesting example is that of the orbit of the Pleiades
binary Atlas. As is well known, the release of the Hipparcos-based
distance to the Pleiades of 118 pc \citep{vL99}, which was some 10\% lower
than the ``canonical'' value of 132 pc from main sequence fitting
\citep{PSS98}, caused a major controversy \citep[e.g.,][]{P04}.
Interestingly, it was the analysis of the visual binary Atlas in the
cluster by \citet*{PSK04} \citep[later refined by][]{ZND04} that opened
the way to the resolution of the problem by obtaining a distance in
agreement with the predictions of stellar models. A recent astrometric
analysis by \citet{SNB05} has indeed revealed a systematic difference
between HST/FGS and Hipparcos parallaxes that could explain the
discrepancy.

These are just a few selected examples of a method with great potential.  
As discussed by \citet{P00} the use of visual binaries to estimate
distances has been neglected in the past. However, the situation is due to
change in the coming years with the launch of missions such as Gaia or SIM
that will push the astrometric limits down to the micro-arcsecond domain.
The prospects for Gaia are especially promising in the determination of
distances to visual binaries because the mission will also obtain
spectroscopic measurements from which radial velocities can be derived.

The second approach to distance determination using binaries relies on its
particularity to yield the fundamental properties of the component stars.  
The procedure is direct and simple but it needs data from different
sources.  The combination of the light and radial velocity curves of
eclipsing binaries yields the orbital and physical properties of the
system. Then, an estimate of the stellar temperatures permits the
calculation of the intrinsic luminosities and the distance follows by
comparison with the observed brightness. There are two caveats worth
mentioning. First, one needs to make sure that no systematic error is
introduced when estimating the effective temperature, which is always
based upon some type of calibration. There are several alternatives to
determine reliable temperatures, such as atmosphere model fits the
observed spectral energy distribution, empirical calibrations based on
photometric indexes, or detailed spectral analyses. Second, interstellar
extinction plays an important role and has to be corrected for.
\citet{C04} provides a general discussion with emphasis on the involved
uncertainties.

It was long realized that eclipsing binaries can be used as powerful
distance indicators \citep{G40,dV78,P96}. However, the method did not
receive much attention until the instrumental capabilities reached
sufficient precision to permit detailed analyses of eclipsing binaries in
the Magellanic Clouds \citep{GFD98}. The number of distance determinations
of Large and Small Magellanic Cloud (LMC and SMC) binaries has increased
in recent years \citep*{FRG02,FRG03,RFM02,HHH03,HHH05}, and more are
expected shortly. Although it may seem that the longstanding problem of
the distance to the LMC and SMC is now resolved \citep{A04}, the scatter
of the distance estimates to individual eclipsing binaries is larger than
the expected error bars. This is discussed by \citet{R04}, who studies the
possibility of a line-of-sight extension of the LMC. It is only with
detailed analyses of further eclipsing binaries that the current issues
can be settled. Extragalactic binary research is a rapidly developing
discipline, with large numbers of new eclipsing binaries in Local Group
galaxies being reported by several surveys. The recent first analyses of
the faint eclipsing binaries in M31 \citep{RJV05} and M33 \citep{Bea05},
requiring the use of the most powerful instruments, constitute a clear
example of the intense activity and interest in the field.

But eclipsing binaries are not only useful as distance indicators to Local
Group galaxies. They can also be employed to estimate accurate distances
to galactic clusters. A particularly relevant example is that of the
Pleiades eclipsing binary HD 23642. Its analysis by \citet{MDS04} and
\citet*{SMS05} has resulted in a distance estimation in excellent
agreement with that from Atlas and from the cluster main sequence fitting.
Many clusters in the Galaxy, which are key objects to our understanding of
stellar evolution and a very important step in the distance ladder, should
have eclipsing binaries that are still awaiting discovery and detailed
analysis.

\subsection{Clocks}

The light-travel time effect (LTTE) in eclipsing binaries produces
periodic variations in the mid-eclipse times with a very simple and direct
physical meaning: The total path that the light has to travel varies
periodically as the eclipsing pair moves around the barycenter of a wider
triple system.  In other words, the eclipses act as an accurate clock for
detecting subtle variations in the distance to the object. This is
analogous to the method used for discovering earth-sized objects around
pulsars \citep{WF92}. The amplitude of the variation is proportional both
to the mass and to the period of the third body, as well as to the sine of
the orbital inclination. The analytical expressions that describe
accurately the LTTE as a function of the orbital properties were first
proposed by \citet{I52}. As discussed by \citet{D00}, nearly 60 eclipsing
binaries show evidence for LTTEs.

Finding additional companions to an eclipsing binary system has limited
interest unless the companion is a special kind of star. For example, the
Timing analysis of the Hyades binary V471 Tau indicates a $\sim$30 yr
modulation with an amplitude of $\sim$140 s. Such values are compatible
with the perturbation from an object with a minimum mass of about
0.04~M$_{\odot}$ \citep{GR01}, which is in the brown dwarf realm. Another
possibility is the combination of astrometric measurements and LTTE to
resolve all the orbital and physical parameters of the third component.
For example, this was done for Algol \citep{BH75} and for R CMa
\citep*{RAG02}. However, this method will reach its full potential with
the upcoming high-accuracy astrometric missions (such as Gaia and SIM).
With timing accuracies of $\sim$10 s for select eclipsing binaries with
sharp eclipses, the detection of large planets ($\sim$10~M$_{\rm J}$) in
long-period orbits ($\sim$10--20~yr) around eclipsing binaries will be a
relatively easy task. The short-term astrometry will confirm the
detections and yield the complete orbital solution and thus the actual
mass of the orbiting body. Finally, transiting planets are also prime
candidates for LTTE studies. In this case, not only further orbiting
planets can be discovered, but even moons around the transiting planet.

\section{Binary star evolution}

Besides providing useful tests to single star evolution models, the
evolution of binary stars (in close, interacting systems) deserves
attention in its own. Mass loss and mass transfer lead to evolutionary
stages and stellar structures that would not be possible otherwise. The
reader is referred to specialized literature on the subject \citep*[][and
references therein]{CH02,dL01,TS00,VDV98}. But also, the orbital evolution
(i.e., circularization and synchronization) of binaries is directly
related to the structure of the components. Orbital circularization is
driven mostly by tidal dissipation. However, the actual dissipative
mechanism in play is the matter of some controversy. Two competing
theories \citep{Z89,T88} predict rather different
circularization and synchronization timescales. Observations are the only
way to test which one of the approaches is indeed physically valid but no
definitive conclusion have been reached yet \citep*{CGC95,CC97}.

Interacting, close binaries are the progenitors of objects with strong
astrophysical interest such as type Ia supernovae, novae, X-ray binaries,
cataclysmic variables, microquasars, symbiotic stars, double degenerates,
etc. Specific reviews on each of these object classes can be found in
\citet{HN00}, \citet{S89}, \citet*{Lvv95}, \citet{W03}, \citet{MR99},
\citet*{CMM03}, \citet{H98}, and references therein. These objects, often
related to energetic phenomena, are the subject of intense study today. An
illustration of this is the strong interest raised by galactic
microquasars, which are scaled-down versions of quasars in which the
accretor is a stellar-mass compact object (neutron star or black hole) and
the donor is a star that loses mass via Roche lobe overflow or stellar
wind \citep{Ri05,MR98}. Besides providing valuable information on
accretion processes at smaller (temporal and spatial) scales than quasars,
microquasars are the source of very high energy emissions
\citep{PMR00,Aea05}, but the mechanism by which such high energies can be
attained has not been pinpointed yet \citep{Ro05,P05}.

To conclude with this short overview of the interest of binaries by
themselves, a point worth mentioning is the foreseen strong impact on
gravitational wave astrophysics. This is a new window to astronomy that
will experience a revolution with the increasing sophistication of the
detectors and the launch of the space mission LISA. Binaries composed of
compact objects such as white dwarfs, neutron stars and black holes, and
the coalescence of the components of the binaries, are expected to be
strong sources of gravitational radiation \citep*{NYP01}. Furthermore, the
shear number of double white dwarfs in the Galaxy is such that their
gravitational wave radiation is anticipated to dominate the background and
even limit the capabilities of the instrumentation \citep{EIS87}.  At this
point, all expectations are based upon theoretical calculations but many
exciting results and new research opportunities are expected for the
coming years with the dawn of observational gravitational wave
astrophysics.

\section{Binaries everywhere}

We are living in the era of the photometric surveys. Many projects have
produced huge amounts of photometric data and more are to come from those
still ongoing. The ground-based projects EROS, MACHO, OGLE, STARE, ASAS,
WASP, ROTSE, and the space missions COROT, Kepler, and Gaia are just a few
examples. The resulting photometric datasets contain a wealth of
information on stellar variability and, from this, many new binaries can
be identified, mostly ellipsoidal and eclipsing binaries. Extensive
catalogs of eclipsing binaries in the Galaxy and the Magellanic Clouds
have already been compiled, greatly increasing (sometimes by orders of
magnitude) the number of known systems.

But some criticisms should be made to hold back the possible euphoria when
facing such bonanza of data. It has to be kept in mind that the majority
of these surveys have not been designed for further exploitation of the
resulting photometry. For example, in the case of eclipsing binaries, the
light curves are usually in a single passband (two at most) and often
undersampled. Also, complementary observations (either photometric or
spectroscopic) may be needed to characterize the binary system. Survey
data is of great use from a statistical point of view to study the
distribution of, e.g., orbital periods, eccentricities, stellar radii,
etc, as shown by, e.g., \citet{NZ03}. But caution must be exercised when
carrying out detailed analyses. Particularly important when dealing with
such large datasets are automatic light curve fitting schemes. Examples of
automatic codes are the recent papers by \citet{WW01,WW02} and
\citet{D05}. Additionally, significant progress needs to be made in the
modeling of fine effects in light curves (gravity brightening, limb
darkening, reflection). Photometry with sub-milimag accuracy from upcoming
space missions will certainly expose the shortcomings of our current
theoretical description of light curves.

\section{Conclusions}

In this paper I have tried to present a brief overview of some of the
areas in which binary stars (may) play an important role. The few aspects
discussed here have been selected to illustrate that binaries can be very
interesting as individuals but, more importantly, they can produce very
valuable contributions to Astrophysics in general. Such astrophysical
insight is central to make a research activity worthwhile. With binary
stars one can address topics so diverse as the cosmic distance scale,
stellar evolution, gravitational waves, which have been discussed here.
But there are also many other aspects, like magnetic activity, plasma
physics, variable stars in binaries, to name a few, that can also be
studied. Some additional examples are given, e.g, in \citet{G93} and the
particular aspect of variable stars in binaries is addressed by Pigulski
and Lampens in this volume. The binary world is rich both in variety 
and in value, and it offers plenty of exciting research opportunities.

\section{Acknowledgments}

I am grateful to the organizers for the invitation to participate in the
PhD School and to the students for their enthusiasm and interest. Support
from the Spanish MEC through a Ram\'on y Cajal fellowship and from the
Spanish MCyT grant AyA2003-07736 is acknowledged.

\end{document}